\documentclass[journal]{IEEEtran}
\usepackage{hyperref}
\usepackage{amssymb,amsmath}
\usepackage{stmaryrd}
\usepackage{tikz}
  \usetikzlibrary{arrows,automata,shapes,backgrounds}
  \usetikzlibrary{decorations.pathmorphing}
  \usetikzlibrary{calc}
\usepackage{subcaption}
\usepackage{array}
\usepackage{cite}
\usepackage{url}
\usepackage{enumitem}
\usepackage[english]{babel}
\newcommand{\eps}{\varepsilon}
\newtheorem{thm}{Theorem}
\newtheorem{cor}[thm]{Corollary}
\newtheorem{lem}[thm]{Lemma}

\newtheorem{defn}[thm]{Definition}

\usepackage[textwidth=1.9cm,color=green!10,textsize=footnotesize]{todonotes}

\begin{document}
\title{Critical Observability for Automata and Petri~Nets}

\author{Tom{\' a}{\v s}~Masopust\thanks{T. Masopust ({\tt masopust{@}math.cas.cz}) is with the Department of Computer Science, Palacky University, Olomouc, Czechia, and with the Institute of Mathematics of the Czech Academy of Sciences, \v Zi\v zkova 22, 616 62 Brno, Czechia. The research was supported by RVO 67985840.}}

\markboth{}{}

\maketitle

\begin{abstract}
  Critical observability is a property of cyber-physical systems to detect whether the current state belongs to a set of critical states. In safety-critical applications, critical states model operations that may be unsafe or of a particular interest. De Santis et al. introduced critical observability for linear switching systems, and Pola et al. adapted it for discrete-event systems, focusing on algorithmic complexity. We study the computational complexity of deciding critical observability for systems modeled as (networks of) finite-state automata and Petri nets. We show that deciding critical observability is (i) NL-complete for finite automata, that is, it is efficiently verifiable on parallel computers, (ii) PSPACE-complete for networks of finite automata, that is, it is very unlikely solvable in polynomial time, and (iii) undecidable for labeled Petri nets, but becoming decidable if the set of critical states (markings) is finite or co-finite, in which case the problem is as hard as the non-reachability problem for Petri nets.
\end{abstract}


\section{Introduction}
  The state estimation problem is one of the central problems in cyber-physical systems that is of importance, e.g., in safety-critical applications where we need to estimate the current state of a system in the case we have an incomplete information of its behavior. Eminent examples of the state estimation problem are, for example, {\em fault diagnosability}~\cite{SantisB17,ZaytoonL13,YinLafortune17} asking whether a fault event has occurred and whether its occurrence can be detected within a finite delay, {\em opacity}~\cite{JacobLF16,TongLSG17,Badouel2007,Lin2011,Bryans2005,SabooriHadjicostis2007,WuLafortune2013}, a property related to the privacy and security analysis, asking whether the system reveals its secret to a passive observer (an intruder), {\em detectability}~\cite{ShuLin2011,ShuLin2013,Zhang17} asking whether the current and subsequent states can be determined based on observations, {\em marking observability}~\cite{GiuaS02} concerning the estimation of the marking of a Petri net, and {\em predictability}~\cite{Fiore2018,GencL09} concerning the future occurrence of a state or of an event.
  
  We study the verification complexity of such a property called {\em critical observability} asking whether the current state of the system, determined based on incomplete observations, is critical. De Santis et al.~\cite{DeSantis2006} introduced the problem for linear switching systems, and Pola et al.~\cite{PolaSBP17} adapted it for (networks of) finite automata. Critical states are of particular interest in safety-critical applications to model operations that may be unsafe or of a specific interest, where, for instance, the prompt recovery of human errors and device failures are of importance to ensure safety of the system, such as the air traffic management systems~\cite{DiBenedetto2005,DeSantis2006,SantisB17}. 
  
  Pola et al. focused on the algorithmic complexity of checking critical observability for systems modeled as networks of finite automata, using the techniques of decentralization and bisimulation. We investigate the computational complexity of this problem for (networks of) finite automata and for labeled Petri nets. Our contributions are as follows.

  We show that the problem of deciding critical observability of finite automata is NL-complete, which means that it can be efficiently verified on a parallel computer~\cite{AroraBarak2009}. Pola et al. showed that critical observability and strong detectability of Shu et al.~\cite{ShuLinYing2007} are different properties. Our result reveals that they are equivalent under the deterministic logarithmic-space reduction~\cite{masopust2018}, that is, critical observability can be reduced to strong detectability by a deterministic algorithm working in logarithmic space, and vice versa. Therefore, any abstraction technique or approximation algorithm for strong detectability can be used for critical observability as well, and vice versa. 
  
  For systems modeled as a network of finite automata, we show that deciding critical observability is PSPACE-complete, and hence there is very unlikely a polynomial-time algorithm solving the problem. 
  
  Finally, we show that critical observability is undecidable for systems modeled by labeled Petri nets, but that it becomes decidable if the set of critical states (markings) is finite or co-finite (a set is co-finite if its complement is finite). We show that, in this case, the problem is as hard as the non-reachability problem for Petri nets. The complexity of reachability for Petri nets has recently been shown to be non-elementary~\cite{WojtekNotElem}.

\section{Preliminaries and Definitions}
  For a set $A$, $|A|$ denotes its cardinality and $2^{A}$ its power set. An alphabet $\Sigma$ is a finite nonempty set of events. A word over $\Sigma$ is a finite sequence of events; $\varepsilon$ denotes the empty word. Let $\Sigma^*$ be the set of all words over $\Sigma$. The alphabet $\Sigma$ is partitioned into two disjoint subsets $\Sigma_o$ of {\em observable\/} and $\Sigma_{uo}=\Sigma\setminus\Sigma_o$ of {\em unobservable\/} events. The partitioning induces a projection $P\colon \Sigma^* \to \Sigma_o^*$, which is a morphism defined by $P(a) = \varepsilon$ for $a\in \Sigma\setminus \Sigma_o$, and $P(a)= a$ for $a\in \Sigma_o$. The action of $P$ on a word $\sigma_1\sigma_2\cdots\sigma_n$ is to erase all events that do not belong to $\Sigma_o$, i.e., $P(\sigma_1\sigma_2\cdots\sigma_n)=P(\sigma_1) P(\sigma_2) \cdots P(\sigma_n)$. 

  We now briefly review the necessary notions of complexity theory and refer the reader to the literature for details~\cite{AroraBarak2009,sipser}.
  A {\em (decision) problem\/} is a yes-no question. A problem is {\em decidable\/} if there is an algorithm that solves it. 
  Complexity theory classifies decidable problems into classes based on time or space an algorithm needs to solve the problem. We consider NL, NP, PSPACE, and EXPSPACE denoting the classes of problems solvable by nondeterministic logarithmic-space, nondeterministic polynomial-time, deterministic polynomial-space, and deterministic exponential-space algorithms, respectively. 
  A problem is NL-complete if it belongs to NL and every problem from NL can be reduced to it in deterministic logarithmic space. Similarly, for $X$ denoting NP, PSPACE, or EXPSPACE, a problem is X-complete if (i) it belongs to X and (ii) every problem from X can be reduced to it in deterministic polynomial time. Condition (i) is known as {\em membership\/} and (ii) as {\em hardness}. 
  By the space hierarchy theorem~\cite{StearnsHL65}, NL is a strict subclass of PSPACE and PSPACE is a strict subclass of EXPSPACE. Moreover, NL is the class of problems efficiently solvable on parallel computers~\cite{AroraBarak2009}. For an EXPSPACE-complete problem, there is neither a polynomial-space nor a polynomial-time algorithm. It is believed that there are no polynomial-time algorithms for NP-complete problems.

\section{Critical Observability for Automata}
  We assume that the reader is familiar with the basic notions and concepts of automata theory~\cite{sipser,Lbook}.

  A {\em nondeterministic finite automaton\/} (NFA) is a quintuple $G = (Q,\Sigma,\delta,I,F)$, where $Q$ is a finite set of states, $I\subseteq Q$ is a nonempty set of initial states, $F \subseteq Q$ is a set of marked states, and $\delta \colon Q\times\Sigma \to 2^Q$ is a transition function that can be extended to the domain $2^Q\times\Sigma^*$ by induction. The {\em language generated by $G$\/} is the set $L(G) = \{w\in \Sigma^* \mid \delta(I,w)\neq\emptyset\}$ and the {\em language marked by $G$\/} is the set $L_m(G) = \{w\in \Sigma^* \mid \delta(I,w)\cap F \neq\emptyset\}$.
  The NFA $G$ is {\em deterministic\/} (DFA) if it has a unique initial state ($|I|=1$) and no nondeterministic transitions ($|\delta(q,a)|\le 1$ for every $q\in Q$ and $a \in \Sigma$). 
  We say that a DFA is {\em total\/} if its transition function is total, that is, $|\delta(q,a)|=1$ for every $q\in Q$ and $a\in\Sigma$.

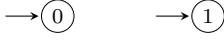
\begin{figure}
    \centering
    \begin{tikzpicture}[baseline,->,>=stealth,auto,shorten >=1pt,node distance=2cm,
      state/.style={circle,minimum size=0mm,inner sep=2pt,very thin,draw=black,initial text=},
      every node/.style={font=\footnotesize}]
      \node[state,initial]  (a) {$0$};
      \node[state,initial]  (aa) [right of=a]  {$1$};
    \end{tikzpicture}
    \caption{Let $G=(\{0,1\},\{a\},\emptyset,\{0,1\},\emptyset)$ be the depicted NFA, $C=\{0\}$. Both states are initial, $0\in C$, $1\notin C$, and $G$ is critically observable with respect to $C$ by the definition of Pola et al.~\cite{PolaSBP17} because $\delta(0,w)\subseteq C$ and $\delta(1,w)\subseteq \{0,1\}\setminus C$ for any $w\in L(G)=\{\eps\}$.}
    \label{fig01}
  \end{figure}

  Given an NFA $G=(Q,\Sigma,\delta,I,F)$ and a set of critical states $C\subseteq Q$. Pola et al.~\cite{PolaSBP17} define $G$ to be critically observable with respect to $C$ if $\delta(i,w) \subseteq C$ or $\delta(i,w) \subseteq Q\setminus C$ for any initial state $i \in I$ and any $w \in L(G)$. 
  They further assume that $I\subseteq C$ or $I\subseteq Q\setminus C$, justifying this assumption by the claim that if $G$ has an initial state that is critical and another initial state that is not critical, then $G$ is never critically observable with respect to $C$. This is misleading as illustrated in Fig.~\ref{fig01}.

  To fix this inconsistency, we can either assume, without loss of generality, that NFAs possess a single initial state, or restate the definition so that $G$ is critically observable with respect to $C$ if $\delta(I,w) \subseteq C$ or $\delta(I,w) \subseteq Q\setminus C$ for any $w \in L(G)$. Then the claim holds and we may assume that $I \subseteq C$ or $I \subseteq Q\setminus C$.
  
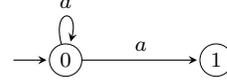
\begin{figure}
    \centering
    \begin{tikzpicture}[baseline,->,>=stealth,auto,shorten >=1pt,node distance=2cm,
      state/.style={circle,minimum size=0mm,inner sep=2pt,very thin,draw=black,initial text=},
      every node/.style={font=\footnotesize}]
      \node[state,initial]  (a) {$0$};
      \node[state]  (aa) [right of=a]  {$1$};
      \path
        (a) edge[loop above] node{$a$} (a)
        (a) edge node{$a$} (aa);
    \end{tikzpicture}
    \caption{An example showing that $I=C$ does not imply that $G$ is critically observable; here $C=I=\{0\}$ and $\delta(0,a)=\{0,1\}$}
    \label{fig02}
  \end{figure}

  Pola et al.~\cite{PolaSBP17} also assume that $I\neq C$ and claim that if $I=C$, then $G$ ''is critically observable and no further analysis for the detection of critical states is needed.`` This claim is again misleading as illustrated in Fig.~\ref{fig02}, and hence we drop this assumption in our paper.

  We further point out that Pola et al.~\cite{PolaSBP17} investigated NFAs with full observation, which includes NFAs with partial observation under the fact that every NFA with partial observation can be transformed to an NFA with full observation. Although it is correct for a single NFA, requiring some computational (polynomial) effort, and hence not suitable for our complexity analysis, it causes serious troubles for networks of automata as we discuss in Section~\ref{section_networks}. Therefore, we do not use this simplification and extend the definition to systems with partial observation.
  
  \begin{defn}
    Let $G=(Q,\Sigma,\delta,I,F)$ be an NFA, $\Sigma_o\subseteq\Sigma$ be the set of observable events, and $C\subseteq Q$ be a set of critical states. Let $P\colon \Sigma^* \to \Sigma_o^*$ denote the induced projection. System $G$ is {\em critically observable\/} with respect to $\Sigma_o$ and $C$ if $\delta(I,P^{-1}P(w)) \subseteq C$ or $\delta(I,P^{-1}P(w)) \subseteq Q\setminus C$ for any $w \in L(G)$, where $\delta(I,P^{-1}P(w))= \cup_{v\in P^{-1}P(w)} \delta(I,v)$. If $P$ is an identity, that is, all events are observable, we simply say that $G$ is critically observable with respect to $C$.
  \end{defn}

\subsection{Single NFA Models}
  We first characterize critical observability in terms of reachability in a composition of two copies of the NFA, and then use this characterization to check critical observability in nondeterministic logarithmic space. Our result reveals that the algorithmic complexity of deciding critical observability is at most quadratic in the number of states.
  
  Let $G=(Q,\Sigma,\delta,I,F)$ be an NFA and $\Sigma_o\subseteq\Sigma$ be the set of observable events. We define a modified parallel composition of two copies of $G$, denoted by $G\interleave G$, as the classical parallel composition where observable events behave as shared events and unobservable events as private events. Formally, $G \interleave G$ is the accessible part of NFA $(Q\times Q, \Sigma, f, I\times I, F\times F)$, where
  \[ f((x,y),e) = 
    \left\{ 
      \begin{array}{l@{\ }l@{}}
        \delta(x,e) \times \delta(y,e) & \text{if } e\in \Sigma_o\\
        (\delta(x,e)\times \{y\}) \cup (\{x\} \times \delta(y,e)) & \text{if } e\notin \Sigma_o
      \end{array}
    \right.
  \]

  Unlike the classical parallel composition~\cite{Lbook}, if $G$ is a DFA, the composition $G\interleave G$ is not necessarily a DFA.

  We now formulate a lemma relating critical observability to reachability in $G\interleave G$.
  \begin{lem}\label{lemma3}
    Let $G=(Q,\Sigma,\delta,I,F)$ be an NFA, $\Sigma_o\subseteq\Sigma$ be the set of observable events, and $C\subseteq Q$ be a set of critical states. Then $G$ is not critically observable with respect to $\Sigma_o$ and $C$ iff there is a reachable state in $G\interleave G$ that belongs to the set $C\times (Q\setminus C)$.
  \end{lem}
  \begin{IEEEproof}
    Let $P$ denote the induced projection from $\Sigma$ to $\Sigma_o$.
    If $G$ is not critically observable, then there are $w\in L(G)$, $x\in C$, and $y\in Q\setminus C$ such that $\{x,y\}\subseteq \delta(I,P^{-1}P(w))$. By the definition of $\interleave$, state $(x,y)$ is reachable in $G\interleave G$. 
    
    For the opposite, we assume that $(x,y)\in C\times (Q\setminus C)$ is reachable in $G\interleave G$. Then, there are $w_1,w_2\in L(G)$ such that $x\in \delta(I,w_1)$, $y\in \delta(I,w_2)$, and $P(w_1)=P(w_2)$. Since $w_1,w_2\in P^{-1}P(w_2)$, we have that $\{x,y\}\subseteq \delta(I,P^{-1}P(w_2))$, which shows that $G$ is not critically observable.
  \end{IEEEproof}

  The use of $G \interleave G$ in Lemma~\ref{lemma3} suggests an algorithm deciding critical observability in time quadratic in the number of states of the NFA. 
  
  We now prove our main result for finite-automata models.
  \begin{thm}\label{thm4}
    Deciding critical observability for systems modeled by NFAs is NL-complete. It remains NL-hard even if the NFAs are with full observation over a unary alphabet and the set of critical states is a singleton.
  \end{thm}
  \begin{IEEEproof}
    Let $G=(Q,\Sigma,\delta,I,F)$ be an NFA and $\Sigma_o\subseteq\Sigma$ be a set of observable events. By Lemma~\ref{lemma3}, $G$ is not critically observable iff there is a reachable state in $G\interleave G$ of the form $C\times (Q\setminus C)$. The nondeterministic algorithm first guesses a state of $C\times (Q\setminus C)$ and then verifies, using the nondeterministic search strategy, that the guessed state is reachable in $G\interleave G$. In this strategy, the algorithm stores only the current state of $G\interleave G$, which in binary requires logarithmic space, and hence the algorithm runs in logarithmic space; cf. the literature for details how to check reachability in NL~\cite{AroraBarak2009,Masopust2018nb}. Thus, deciding whether $G$ is not critically observable belongs to NL. Since NL is closed under complement~\cite{Immerman88,Szelepcsenyi87}, deciding critical observability belongs to NL as well.
    
    \begin{figure}
      \centering
      \begin{tikzpicture}[baseline,auto,->,>=stealth,shorten >=1pt,node distance=1.3cm,
        state/.style={ellipse,minimum size=5mm,inner sep=2pt,very thin,draw=black,initial text=},
        every node/.style={font=\small}]
        \node[state,initial]  (1) {$s$};
        \node[state]          (2) [above of=1,node distance=.8cm]  {$p$};
        \node[state]          (5) [right of=2,label={right:$D$}]  {$q$};
        \node[state]          (3) [right of=1,node distance=2.7cm]  {$t$};
        \node[state]          (4) [right of=3,node distance=1.7cm]  {$r$};
        \path
          (2) edge node[sloped,above] {$a$} (5)
          (3) edge node[pos=0.5,sloped,above] {$a$} (4)
          (3) edge[loop above] node {$a$} (3)
          (1) edge[style={decorate, decoration={snake,amplitude=.4mm,segment length=1.7mm,post length=1.3mm}}] node{?} (3) ;
          ;
        \begin{pgfonlayer}{background}
          \path (2.north -| 3.east) + (0.1,0.1)    node (a) {};
          \path (1.south -| 1.west) + (-0.3,-0.1)  node (b) {};
          \path[rounded corners, draw=black] (a) rectangle (b);
        \end{pgfonlayer}
      \end{tikzpicture}
      \caption{The NFA $G$ of the NL-hardness proof of Theorem~\ref{thm4}}
      \label{fig1}
    \end{figure}
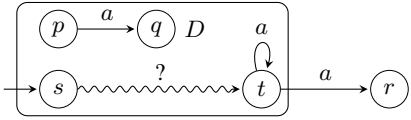

    To show NL-hardness, we reduce {\em DAG non-reachability}~\cite{ChoH91}: Given a directed acyclic graph $D=(V,E)$ and nodes $s,t\in V$, is $t$ not reachable from $s$? From $D$, we construct the NFA $G=(Q,\{a\},\delta,s,Q)$, where $Q=V\cup\{r\}$, $r\notin V$ is a new state, and $a$ is an observable event. For every $(p,q)\in E$, we add the transition $(p,a,q)$ to $\delta$. Then, we add the transitions $(t,a,t)$ and $(t,a,r)$ to $\delta$, cf. Fig.~\ref{fig1}. We show that $t$ is not reachable from $s$ in $D$ iff $G$ is critically observable with respect to $\{a\}$ and $\{t\}$. If $t$ is not reachable from $s$ in $D$, then, for every $w\in\{a\}^*$, $t\notin \delta(s,w)$, which means that $\delta(s,w) \subseteq Q\setminus \{t\}$, and hence $G$ is critically observable. If $t$ is reachable from $s$, let $w$ be such that $t\in \delta(s,w)$. Then $\{t,r\} \subseteq \delta(s,wa)$, and hence $G$ is not critically observable.
  \end{IEEEproof}
  
  If, in the NL-hardness proof, we do not add the transition $(t,a,t)$, replace the transition $(t,a,r)$ by a transition $(t,u,r)$, where $u$ is unobservable, and label every other transition with a fresh new observable event, then the construction results in a DFA and we have the following corollary.
  
  \begin{cor}
    Deciding critical observability for DFAs is NL-complete even if the DFA has a single unobservable event and the set of critical states is a singleton.
    \hfill\IEEEQED
  \end{cor}

  The unobservable event is unavoidable because any DFA with full observation is always in a unique state, and hence trivially critically observable.

\subsection{Networks of Automata Models}\label{section_networks}
  Large-scale systems are often modeled as a composition of local modules $\{G_1,G_2,\dots,G_n\}$ for $n\ge 2$, where $G_i$ is an NFA, i.e., the overall system behaves as $G_1\parallel G_2\parallel \cdots\parallel G_n$. We call such a system a {\em network of NFAs\/}.

  Pola et al.~\cite{PolaSBP17} used observers for checking critical observability. They showed that a decentralized observer for networks of NFAs is isomorphic (denoted by $\approx$) to the composition of local observers. In other words, they showed that
  \begin{align}\label{polaeq}
    Obs(G_1 \parallel \ldots \parallel G_n) \approx Obs(G_1) \parallel \ldots \parallel Obs(G_n)
  \end{align}
  where $Obs(G)$ denotes the observer of $G$~\cite{Lbook,PolaSBP17}. This leads to the decrease of complexity, and Pola et al.~\cite[Table~1]{PolaSBP17} showed that the algorithmic complexity of deciding critical observability for networks of NFAs with full observations is single exponential in time and space.

  To explain why \eqref{polaeq} holds in their setting, notice that they use NFAs with full observations. Therefore, the computation of the observer reduces to the determinisation of an NFA.
  
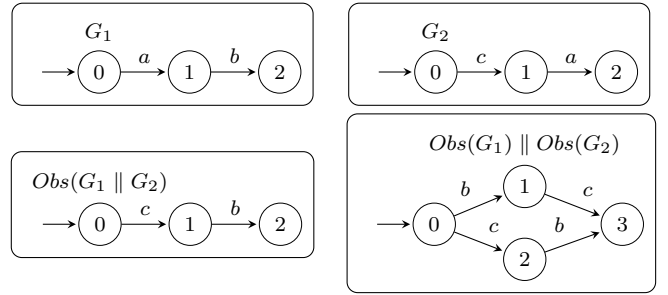
\begin{figure}
    \begin{subfigure}[t]{.24\textwidth}
      \centering
      \begin{tikzpicture}[baseline,auto,->,>=stealth,shorten >=1pt,node distance=1.2cm,
      state/.style={circle,minimum size=0mm,very thin,draw=black,initial text=},
      every node/.style={font=\footnotesize},framed, rounded corners]
      \node[state,initial]            (1) [label={[anchor=south]above:$G_1$}] {$0$};
      \node[state]                    (2) [right of=1] {$1$};
      \node[state]                    (3) [right of=2] {$2$};
      \path
        (1) edge node {$a$} (2)
        (2) edge node {$b$} (3)
        ;
      \end{tikzpicture}
    \end{subfigure}
    \begin{subfigure}[t]{.24\textwidth}
      \centering
      \begin{tikzpicture}[baseline,auto,->,>=stealth,shorten >=1pt,node distance=1.2cm,
        state/.style={circle,minimum size=0mm,very thin,draw=black,initial text=},
        every node/.style={font=\footnotesize},framed, rounded corners]
        \node[state,initial]            (1) [label={above:$G_2$}] {$0$};
        \node[state]                    (2) [right of=1] {$1$};
        \node[state]                    (3) [right of=2] {$2$};
        \path
          (1) edge node {$c$} (2)
          (2) edge node {$a$} (3)
          ;
      \end{tikzpicture}
    \end{subfigure}

    \vspace{.1cm}
    \begin{subfigure}[t]{0.24\textwidth}
      \centering
      \begin{tikzpicture}[baseline,auto,->,>=stealth,shorten >=1pt,node distance=1.2cm,
        state/.style={circle,minimum size=0mm,very thin,draw=black,initial text=},
        every node/.style={font=\footnotesize},framed, rounded corners]
        \node[state,initial]            (1) [label={[anchor=south]above:$Obs(G_1\parallel G_2)$}] {$0$};
        \node[state]                    (2) [right of=1] {$1$};
        \node[state]                    (3) [right of=2] {$2$};
        \path
          (1) edge node {$c$} (2)
          (2) edge node {$b$} (3)
          ;
      \end{tikzpicture}
    \end{subfigure}
    \hfill
    \begin{subfigure}[t]{0.24\textwidth}
      \centering
      \begin{tikzpicture}[baseline,auto,->,>=stealth,shorten >=1pt,node distance=1.6cm,
        state/.style={circle,minimum size=0mm,very thin,draw=black,initial text=},
        every node/.style={font=\footnotesize},framed, rounded corners]
        \node[state,initial]            (1) {$0$};
        \node[state]                    (2) [label={above:$Obs(G_1)\parallel Obs(G_2)$},above right=.1cm and .8cm of 1] {$1$};
        \node[state]                    (4) [right of=1,node distance=2.5cm] {$3$};
        \node[state]                    (3) [below=.4cm of 2] {$2$};
        \path
          (1) edge node {$b$} (2)
          (1) edge node {$c$} (3)
          (2) edge node {$c$} (4)
          (3) edge node {$b$} (4)
          ;
      \end{tikzpicture}
    \end{subfigure}
    \caption{The NFAs $G_1$, $G_2$, $Obs(G_1\| G_2)$, and $Obs(G_1)\parallel Obs(G_2)$; event $a$ is unobservable, $b$ and $c$ are observable}
    \label{fig03}
  \end{figure}

  However, it is known that \eqref{polaeq} does not hold for networks of NFAs with partial observation. This is equivalent to the fact that for two languages $L_1$ and $L_2$ and a projection $P$, we only have $P(L_1\parallel L_2) \subseteq P(L_1) \parallel P(L_2)$~\cite{wm17}. Therefore, for networks of NFAs with partial observation, considering only NFAs with full observation as in the settings of Pola et al.~\cite{PolaSBP17} oversimplifies the situation as illustrated in Fig.~\ref{fig03}.
  
  Considering automata with partial observation may make the complexity infeasible. For instance, we have shown that deciding detectability, opacity, and A-diagnosability for networks of automata is EXPSPACE-complete~\cite{MasopustYin2017arxiv}. The space hierarchy theorem~\cite{StearnsHL65} then implies that there is neither a polynomial-time nor a polynomial-space algorithm for checking these problems for networks of NFAs. 
  
  However, we show that deciding critical observability for networks of NFAs is PSPACE-complete, and hence solvable in polynomial space. Our result thus further generalizes and improves the results of Pola et al.~\cite[Table~1]{PolaSBP17}, who suggested algorithms that are exponential with respect to both time and space.
  \begin{thm}\label{thm6}
    Deciding critical observability for networks of NFAs is PSPACE-complete. It remains PSPACE-hard even if the automata are binary NFAs with full observation, and the set of critical states is a singleton.
  \end{thm}
  \begin{IEEEproof}
    Let $\{G_1,G_2,\ldots,G_n\}$ be a network of NFAs where $G_i=(Q_i,\Sigma_i,\delta_i,I_i,F_i)$, and let $C\subseteq Q_1\times\cdots\times Q_n$ be a set of critical states. A nondeterministic polynomial-space algorithm deciding critical observability generalizes that of Theorem~\ref{thm4} for a single NFA; namely, we consider $G\interleave G$, where $G= \|_{i=1}^{n} G_i$, and the nondeterministic algorithm keeps track of the current state of $G\interleave G$, which in binary requires a polynomial space to store the two $n$-tuples of states, without computing $G$ and $G\interleave G$. Again, the algorithm guesses a state of $C\times (Q\setminus C)$ and uses the nondeterministic search strategy to check that the guessed state is reachable. Since NPSPACE and PSPACE coincide~\cite{Savitch70} and PSPACE is closed under complement, deciding critical observability is in PSPACE.
    
    To show PSPACE-hardness, we reduce the DFA intersection problem~\cite{Kozen77}: Given total DFAs $A_1,\ldots, A_n$ over $\{0,1\}$, is $\cap_{i=1}^{n} L_m(A_i) = \emptyset$? From $A_i=(Q_i,\{0,1\},\delta_i,q_0^i,F_i)$, we construct another DFA $G_i=(Q_i \cup\{s_i\},\{0,1\},\delta_i,q_0^i,F_i)$ by adding a new state $s_i$ and the transition $(p,1,s_i)$ to $\delta_i$ for every $p\in F_i$. We show that $\cap_{i=1}^{n} L_m(A_i)=\emptyset$ iff $G=\|_{i=1}^{n} G_i$ is critically observable with respect to the set $\{(s_1,\ldots,s_n)\}$. If $w\in \cap_{i=1}^{n} L_m(A_i)$, then $s_i\in \delta_i(q_0^i,w1)$, and since $A_i$ is total, $|\delta_i(q_0^i,w1)|\ge 2$, for $i=1,\ldots,n$. Hence $G$ is not critically observable.
    If $\cap_{i=1}^{n} L_m(A_i)=\emptyset$, then $G$ never reaches a marked state, neither the critical state, and hence $G$ is critically observable.
  \end{IEEEproof}

  Actually, PSPACE-hardness holds even if the automata are ternary DFAs with a single unobservable event.
  \begin{cor}\label{cor7}
    Deciding critical observability for networks of automata is PSPACE-complete even if the automata are DFAs with three events, one of which is unobservable, and the set of critical states is a singleton.
  \end{cor}
  \begin{IEEEproof}
    Membership in PSPACE was shown above. To show PSPACE-hardness, we reduce the DFA intersection problem. Let $A_1,\ldots, A_n$ be total DFAs over $\{0,1\}$. From $A_i=(Q_i,\{0,1\},\delta_i,q_0^i,F_i)$, we construct another DFA $G_i=(Q_i \cup\{s_i\},\{0,1,u\},\delta_i,q_0^i,F_i)$ by adding a new state $s_i$ and the transition $(p,u,s_i)$ to $\delta_i$ for every $p\in F_i$. We show that $\cap_{i=1}^{n} L_m(A_i)=\emptyset$ iff $G=\|_{i=1}^{n} G_i$ is critically observable with respect to $\{(s_1,\ldots,s_n)\}$. If $w\in \cap_{i=1}^{n} L_m(A_i)$, then $s_i\in \delta_i(q_0^i,P^{-1}P(wu))$ and $|\delta_i(q_0^i,P^{-1}P(wu))|\ge 2$, for $i=1,\ldots,n$, because $w\in P^{-1}P(wu)$ and $A_i$ is total, and hence $G$ is not critically observable.
    If $\cap_{i=1}^{n} L_m(A_i)=\emptyset$, then $G$ never reaches a marked state, neither the critical state, implying that $G$ is critically observable.
  \end{IEEEproof}

  The unobservable event used in the previous corollary is unavoidable because any network of DFAs with all events observable is always in a unique state, and hence trivially critically observable. 
  
  We now show that two observable events used in Theorem~\ref{thm6} and Corollary~\ref{cor7} are necessary to obtain PSPACE-hardness. As shown in Theorem~\ref{thm7} below, having a single observable event decreases the complexity of the problem.
  \begin{thm}\label{thm7}
    Deciding critical observability for a network of unary NFAs is coNP-complete.
  \end{thm}
  \begin{IEEEproof}
    Since coNP is the class of problems the complement of which belongs to NP, we prove the claim by showing that the problem whether a system is {\em not\/} critically observable is NP-complete.
  
    To show membership in NP, assume that the system consists of $n$ unary NFAs, each of which has at most $k$ states. Then the parallel composition of the NFAs has at most $2^{kn}$ states. If the system is not critically observable, then there is $0\le\ell\le 2^{kn}$ such that $0^{\ell}$ leads the system to two states one of which is critical and the other is not. A nondeterministic polynomial-time algorithm can guess $\ell$ in binary, which is of polynomial length $O(kn)$, and use the matrix multiplication to compute the set of states reachable under $0^{\ell}$ in polynomial time; cf. the literature for details how to use matrix multiplication~\cite{Masopust2018nb}. Having this set of states, it is easy to verify whether the guess was correct.
    
    To show that the problem whether a system is not critically observable is NP-hard, we reduce the nonempty intersection problem for unary NFAs~\cite{StockmeyerM73}. Let $A_1,\ldots,A_n$ be NFAs over a unary alphabet $\{a\}$. From $A_i=(Q_i,\{a\},\delta_i,I_i,F_i)$, we construct an NFA $G_i=(Q_i\cup\{s_i,t_i\},\{a\},\delta_i,I_i,F_i)$ by adding two new states $s_i$ and $t_i$, and transitions $(p,a,s_i)$ and $(p,a,t_i)$, for every $p\in F_i$. Then $\|_{i=1}^{n} G_i$ is not critically observable with respect to $\{(s_1,\ldots,s_n)\}$ iff $\cap_{i=1}^{n} L_m(A_i)$ is nonempty.  
  \end{IEEEproof}

\section{Critical Observability for Petri Nets}
  We assume that the reader is familiar with the basic notions and concepts of Petri nets~\cite{Peterson1981}. Let $\mathbb{N}$ denote the set of all natural numbers (including zero).

  A {\em Petri net\/} is a structure $N=(P,T,Pre,Post)$, where $P$ is a finite set of places, $T$ is a finite set of transitions, $P \cup T \neq \emptyset$ and $P \cap T = \emptyset$, and $Pre\colon P \times T \to \mathbb{N}$ and $Post\colon P \times T \to \mathbb{N}$ are the pre- and post-incidence functions specifying the arcs directed from places to transitions and vice versa. A marking is a function $M\colon P \to \mathbb{N}$ assigning to each place a number of tokens. A Petri net system $(N, M_0)$ is the Petri net $N$ with the initial marking $M_0$.
  A transition $t$ is enabled in a marking $M$ if $M(p) \ge Pre(p,t)$ for every place $p\in P$. If $t$ is enabled, it can fire, resulting in the marking $M(p) - Pre(p,t) + Post(p,t)$ for every $p\in P$. Let $M \xrightarrow{\sigma}_{N}$ denote that the transition sequence $\sigma$ is enabled in marking $M$ of $N$, and $M \xrightarrow{\sigma}_N M'$ that the firing of $\sigma$ results in a marking $M'$. We often omit the subscript $N$ if it is clear from the context. Let $L(N,M_0) = \{ \sigma \in T^* \mid M_0\xrightarrow{\sigma}\}$ denote the set of all transition sequences enabled in marking $M_0$.

  A {\em labeled Petri net system\/} is a structure $G=(N,M_0,\Sigma,\ell)$, where $(N,M_0)$ is a Petri net system, $\Sigma$ is an alphabet (a set of labels), and $\ell\colon T \to \Sigma\cup\{\eps\}$ is a labeling function that can be extended to $\ell\colon T^* \to \Sigma^*$ by $\ell(\sigma t) = \ell(\sigma)\ell(t)$ for $\sigma \in T^*$ and $t \in T$; we set $\ell(\lambda) = \eps$ for $\lambda$ denoting the empty transition sequence. A transition $t$ is observable if $\ell(t)\in\Sigma$ and unobservable otherwise. The language of $G$ is the set $L(G) = \{ \ell(\sigma) \mid \sigma \in L(N,M_0) \}$. 
  A marking $M$ is reachable in $G$ if there is a sequence $\sigma \in T^*$ such that $M_0 \xrightarrow{\sigma} M$. The set of all markings reachable from the initial marking $M_0$ defines the reachability set of $G$, denoted by $R(G)$. For $s\in L(G)$, let $R(G,s) = \{ M \mid \sigma\in L(N,M_0),\, \ell(\sigma)=s,\, M_0\xrightarrow{\sigma}M \}$ be the set of all reachable markings consistent with the observation~$s$.

  \begin{defn}
    Let $G=(N,M_0,\Sigma,\ell)$ be a labeled Petri net, and let $C$ be a set of critical markings. System $G$ is {\em critically observable\/} with respect to $C$ if $R(G,w) \subseteq C$ or $R(G,w) \subseteq R(G)\setminus C$ for every $w \in L(G)$.
  \end{defn}

\subsection{Results}
  Similarly as for automata, checking critical observability is equivalent to checking whether there are two sequences with the same observations leading to two different markings one of which is critical. To formalize this claim, we adopt the twin-plant construction for Petri nets used to test diagnosability~\cite{cabasino2012new,yin2017decidability} or prognosability~\cite{yin2018prognosis}.

  For a labeled Petri net system $G=(N,M_0,\Sigma,\ell)$, let $G'=(N',M_0',\Sigma,\ell)$ be a place-disjoint copy of $G$, that is, $N'=(P',T,Pre',Post')$ where $P'=\{p' \mid p\in P\}$ is a disjoint copy of $P$ and the functions $Pre'$ and $Post'$ are naturally adjusted. The copy $G'$ has the same initial marking as $G$, i.e., $M_0'(p')=M_0(p)$ for every $p \in P$.
 
  Let $(N_{\|},M_{0,\|})=((P_{\|},T_{\|},Pre_{\|},Post_{\|}),M_{0,\|})$ denote a label-based synchronization of $G$ and $G'$, where
  the initial marking $M_{0,\|} = \{M_0\} \times \{M_0'\}$ is the concatenation of initial markings of $G$ and $G'$,
  $P_{\|}=P\cup P'$,
  $T_{\|}= (T\cup\{\lambda\})\times (T\cup\{\lambda\})\setminus \{(\lambda,\lambda)\}$ are pairs of transitions of $G$ and $G'$ without the empty pair, and
  $Pre_{\|}\colon P_{\|}\times T_{\|}\to \mathbb{N}$ and $Post_{\|}\colon P_{\|}\times T_{\|}\to \mathbb{N}$ are defined as follows:
    For $p\in P$ and $t\in T$ with $\ell(t)=\eps$, 
      $Pre_{\|}( p,(t,\lambda) )= Pre(p,t)$, $Post_{\|}( p,(t,\lambda) )= Post(p,t)$,
      $Pre_{\|}( p',(\lambda,t) )= Pre'(p',t)$, $Post_{\|}( p',(\lambda,t) )= Post'(p',t)$,
    and for $p\in P$ and $t_1,t_2\in T$ with $\ell(t_1)=\ell(t_2)\neq\eps$, 
      $Pre_{\|}( p, (t_1,t_2) ) = Pre(p,t_1)$, $Post_{\|}( p, (t_1,t_2) )=Post(p,t_1)$, and
      $Pre_{\|}( p', (t_1,t_2) ) = Pre'(p',t_2)$, $Post_{\|}( p', (t_1,t_2) ) = Post'(p',t_2)$.
    Otherwise, $Pre_{\|}(p_{\|},t_{\|})=Pre_{\|}(p_{\|},t_{\|})=0$, i.e., no arc is defined.

  Intuitively, $(N_{\|},M_{0,\|})$ tracks all pairs of sequences with the same observation; namely, for any $(\sigma,\sigma')\in L(N_{\|},M_{0,\|})$, we have $\ell(\sigma)=\ell(\sigma')$, and for any $\sigma,\sigma'\in L(N,M_0)$ with $\ell(\sigma)=\ell(\sigma')$, there is a sequence in $(N_{\|},M_{0,\|})$ whose first and second components are $\sigma$ and $\sigma'$, respectively (possibly with inserted empty transition $\lambda$).

  The following lemma shows how to use $(N_{\|},M_{0,\|})$ to verify critical observability.
 \begin{lem}\label{thm_formula}
    A labeled Petri net $G=(N,M_0,\Sigma,\ell)$ is not critically observable iff there is a reachable marking $M$ in $(N_{\|},M_{0,\|})$ such that $M(P)\in C$ and $M(P')\notin C$, where $M(P)$ and $M(P')$ are projections of $M$ to the places of $P$ and $P'$, respectively.
  \end{lem}
  \begin{IEEEproof}
    If there is such a reachable marking $M$, then there is a transition sequence $(\alpha,\beta)$ in $L(N_{\|},M_{0,\|})$ such that $M_{0,\|} \xrightarrow{(\alpha,\beta)}_{N_{\|}} M$. By the definition of $(N_{\|},M_{0,\|})$, we have that $M_0 \xrightarrow{\alpha}_N M(P)$, $M_0 \xrightarrow{\beta}_N M(P')$, and $\ell(\alpha) = \ell(\beta)$. Since $M(P) \in C$, $M(P')\notin C$, and $\{M(P),M(P')\}\subseteq R(G,\ell(\alpha))$, $G$ is not critically observable with respect to $C$.

    Assume that the system is not critically observable. Then there is a word $w$ such that $R(G,w)\cap C \neq \emptyset \neq R(G,w)\cap (R(G)\setminus C)$. Let $\alpha,\beta \in L(N,M_0)$ be such that $\ell(\alpha) = \ell(\beta) = w$, $M_0\xrightarrow{\alpha}_N  M_{\alpha} \in C$, and $M_0 \xrightarrow{\beta}_N  M_{\beta} \notin C$. By the construction of $N_{\|}$, we have that $(\alpha,\beta)\in L(N_{\|},M_{0,\|})$, and hence $M_{0,\|} \xrightarrow{(\alpha,\beta)}_{N_{\|}} M=[M_{\alpha} \ M_{\beta}]$, as required.
  \end{IEEEproof}

  To prove our next result, we recall a fragment of Yen's path logic, for which the satisfiability problem is decidable~\cite{yen1992unified,AtigH11}. Let $(N,M_0)$ be a Petri net. Let $M_1,M_2,\ldots$ be variables representing markings and $\sigma_1,\sigma_2,\ldots$ be variables representing finite sequences of transitions. Terms are defined as follows. Every mapping $c \in \mathbb{N}^P$ is a term. For all $j > i$, if $M_i$ and $M_j$ are marking variables, then $M_j - M_i$ is a term, and if $T_1$ and $T_2$ are terms, then $T_1+T_2$ and $T_1-T_2$ are terms. 
  If 
    $c \in \mathbb{N}$
    and $t \in T$, 
  then 
    $\#_t(\sigma_1) \le c$ is an atomic (transition) predicate denoting the number of occurrences of $t$ in $\sigma_1$.
  If $T_1$ and $T_2$ are terms and $p_1,p_2 \in P$ are places, 
  then 
    $T_1(p_1) = T_2(p_2)$, $T_1(p_1) < T_2(p_2)$, and $T_1(p_1) > T_2(p_2)$ are atomic (marking) predicates.
  A {\em predicate\/} is a positive finite boolean combination of atomic predicates. A {\em path formula\/} is a formula of the form
  $
    (\exists \sigma_1, \sigma_2,\ldots, \sigma_n)
    (\exists M_1,\ldots, M_n)
    (M_0 \xrightarrow{\sigma_1} M_1 \xrightarrow{\sigma_2} \cdots \xrightarrow{\sigma_n} M_n) 
    \land
    \varphi(M_1,\ldots,M_n,\sigma_1,\ldots,\sigma_n)
  $
  where $\varphi$ is a predicate.

  We can now prove the following.
  \begin{thm}\label{thm10}
    If the set of critical markings is finite, then critical observability for labeled Petri nets is decidable. It is as hard as the non-reachability problem for Petri nets.
  \end{thm}
  \begin{IEEEproof}
    According to Lemma~\ref{thm_formula}, a labeled Petri net $G=(N,M_0,\Sigma,\ell)$ is not critically observable iff the following path formula of Yen's logic is satisfiable:
    \begin{multline*}
      (\exists \sigma_1,\sigma_2)
      (\exists M_1, M_2)
      (M_{0,\|} \xrightarrow{\sigma_1}_{N_{\|}} M_1 \xrightarrow{\sigma_2}_{N_{\|}} M_2)
      \land \\
      \bigvee_{c \in C} {M_2(P)=c} \land \bigwedge_{c \in C} {M_2(P')\neq c} 
      \land \sigma_1=\eps
    \end{multline*}
    where $\sigma_1=\eps \equiv \land_{t\in T} \#_t(\sigma_1) \le 0$, for $c=(c_i)_{i=1}^{|P|}$, $M_2(P)=c \equiv \land_{i=1}^{|P|} M_2(p_i)=c_i$, $M_2(P)\neq c \equiv \lor_{i=1}^{|P|} (M_2(p_i) < c_i \lor M_2(p_i) > c_i)$, and $M_2 \equiv (M_2-M_1)+M_0$ is a term. 
    Satisfiability of Yen's logic is polynomially reducible to the reachability problem for Petri nets~\cite{AtigH11,yen1992unified}, and hence so is the problem whether $G$ is not critically observable.
    
    We now reduce the reachability problem to the problem of non-critical observability. Let $(N,M_0)$ be a Petri net and $M$ be a marking. We construct a labeled Petri net $G$ by adding a new place $p'$ and a new unobservable transition $t'$ with an arc from $t'$ to $p'$ generating an arbitrary number of tokens in $p'$, that is, $Pre(p',t')=0$ and $Post(p',t')=1$, and with the labeling function $\ell\colon T\cup\{t'\}\to T\cup\{\eps\}$ defined by $\ell(t)=t$, for $t\in T$, and $\ell(t')=\eps$. Let the set of critical markings be $C=\{M\times (0)\}$, where $M\times (0)$ denotes the marking of the net $G$ that coincides with the marking $M$ on the places of the net $N$ and has zero tokens in the new place $p'$. The initial marking of $G$ is the marking $M_0\times (0)$.  Now, if $M$ is not reachable in $(N,M_0)$, then $M\times (0)$ is not reachable in $G$, and hence $G$ is critically observable. However, if $M$ is reachable in $(N,M_0)$, let $\sigma$ denote a transition sequence reaching $M$ in $(N,M_0)$. Then, by construction, $\{M\times (0),M\times (1), M\times (2),\ldots\}\subseteq R(G,\ell(\sigma))$ are reachable in $G$ under sequences with the same labels, since $\ell(\sigma) = \ell(\sigma t') = \ell(\sigma t't') = \ldots$, and hence $G$ is not critically observable.
  \end{IEEEproof}

  The complexity of reachability for Petri nets is a longstanding open problem. The lower bound has recently been improved from EXPSPACE-hard to non-elementary~\cite{WojtekNotElem}. The upper bound is non-primitive recursive cubic Ackermannian~\cite{LerouxS15}. 
  
  We have shown that critical observability is decidable for a labeled Petri net system $G$ if the set of critical markings $C$ is finite. The same holds if the set $R(G)\setminus C$ is finite, which can be shown by exchanging the sets $C$ and $R(G)\setminus C$. If $R(G)\setminus C$ is finite, then $C$ is called co-finite. 
  
\begin{figure}
    \centering
    \includegraphics[scale=.83]{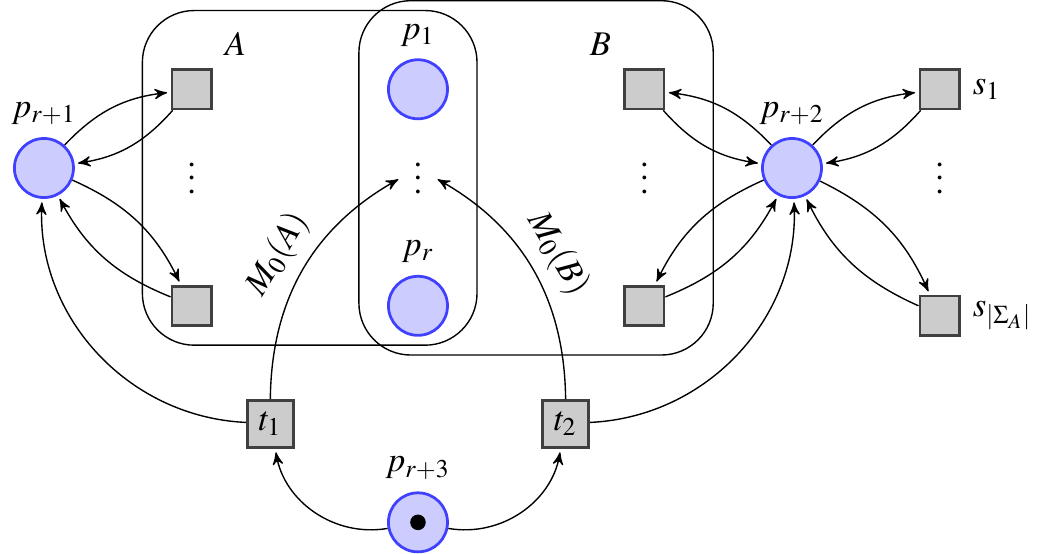}
    \caption{The Petri net $G$ of Theorem~\ref{thm11}}
    \label{fig04}
  \end{figure}

  However, if $C$ is not finite neither co-finite, we show that the problem of critical observability is undecidable.
  \begin{thm}\label{thm11}
    Critical observability for labeled Petri Nets is undecidable.
  \end{thm}
  \begin{IEEEproof}
    We reduce the marking inclusion problem asking, given two Petri nets $A$ and $B$, whether $R(A) \subseteq R(B)$~\cite{Hack76}. 
    Let $\ell_A$ and $\ell_B$ be arbitrary labeling functions of $A$ and $B$, and let $\Sigma_A$ and $\Sigma_B$ denote the corresponding sets of labels. We construct a Petri net $G$ as depicted in Fig.~\ref{fig04}, where place $p_{r+2}$ contains $|\Sigma_A|$ self-loops under new transitions $s_1,\ldots,s_{|\Sigma_A|}$. The initial marking of $G$ consists of one token in place $p_{r+3}$. 
    Then, $R(G) = \{0\}^r\times (0,0,1) \cup R(A)\times (1,0,0) \cup R(B)\times (0,1,0)$. We define the labeling function $\ell$ of $G$ as the extension of $\ell_A$ and $\ell_B$ so that $\ell(t)=\ell_X(t)$ if $t$ is a transition of $X\in\{A,B\}$, $\ell(t_1)=\ell(t_2)=\eps$, and the self-loops in $p_{r+2}$ are labeled by $\Sigma_A$ in such a way that $\ell(s_i)$ is the $i$-th element of $\Sigma_A$. 
    Let $C = \{0\}^r \times (0,0,1) \cup R(B)\times (1,0,0) \cup R(B) \times (0,1,0)$ be the set of critical markings. 
    If $R(A)\subseteq R(B)$, then we have that $R(G)\subseteq C$, and hence $G$ is critically observable. 
    However, if there is a marking $M\in R(A)\setminus R(B)$, let $\sigma$ denote a transition sequence under which $M$ is reachable in $A$. Then the marking $M\times (1,0,0)$ is reachable in $G$ by $t_1\sigma$. Let $\sigma'\in\{s_1,\ldots,s_{|\Sigma_A|}\}^*$ be a sequence of transitions such that $\ell(\sigma)=\ell(\sigma')$; such a sequence exists by the labeling of these places. Then $M_0(B) \times (0,1,0)$ is reachable in $G$ by $t_2\sigma'$. Since $\ell(t_1\sigma)=\ell(t_2\sigma')$, $M\times (1,0,0)\notin C$, and $M_0(B) \times (0,1,0)\in C$, $G$ is not critically observable.
  \end{IEEEproof}
  
\bibliographystyle{IEEEtranS}
\bibliography{mybib}

\end{document}